\documentclass[sigconf,natbib=true,anonymous=false]{acmart}

\graphicspath{{./images/}}

\usepackage{pifont}

\usepackage{multirow}
\usepackage{multicol}
\usepackage{subfigure}
\usepackage{bm}
\usepackage{color}
\usepackage{adjustbox}
\usepackage{balance}

\usepackage{enumitem}
\usepackage{bbm}
\usepackage{graphicx}
\usepackage{mathtools}
\usepackage{bbm}
\usepackage{balance}
\usepackage{graphics}
\usepackage{bbding}
\usepackage{algorithm}
\usepackage{algpseudocode}
\usepackage{booktabs} 
\AtBeginDocument{%
  \providecommand\BibTeX{{%
    \normalfont B\kern-0.5em{\scshape i\kern-0.25em b}\kern-0.8em\TeX}}}

\setcopyright{none}

\copyrightyear{2023} 
\acmYear{2023} 
\acmConference[CIKM '23]{Proceedings of the 32nd ACM International Conference on Information and Knowledge Management}{October 21--25, 2023}{Birmingham, United Kingdom} 
\acmBooktitle{The 1st workshop on recommendation with generative models, October 21--25, 2023, Birmingham, United Kingdom} 
\begin{document}

\title{Is ChatGPT a Good Recommender?  A Preliminary Study}
\author{Junling Liu}
\authornote{Both authors contributed equally to this research.}
\email{william.liuj@gmail.com}
\affiliation{%
  \institution{Alibaba Group}
  \country{China}
}
\author{Chao Liu}
\authornotemark[1]
\email{chize.lc@antgroup.com}
\affiliation{%
  \institution{Ant Group}
  \country{China}
}

\author{Peilin Zhou}
\authornotemark[1]
\email{zhoupalin@gmail.com}
\affiliation{%
  \institution{Hong Kong University of Science and Technology(Guangzhou)}
  \country{China}
}

\author{Renjie Lv}
\email{lvrenjie.lrj@antgroup.com}
\affiliation{%
  \institution{Ant Group}
  \country{China}
}

\author{Kang Zhou}
\email{kangbeyond89@163.com}
\affiliation{%
  \institution{Alibaba Group}
  \country{China}
}

\author{Yan Zhang}
\email{yanbest0117@163.com}
\affiliation{%
  \institution{Alibaba Group}
  \country{China}
  }

\renewcommand{\shortauthors}{Junling Liu, et al.}

\begin{abstract}
Recommendation systems have witnessed significant advancements and have been widely used over the past decades. However, most traditional recommendation methods are task-specific and therefore lack efficient generalization ability. Recently, the emergence of ChatGPT has significantly advanced NLP tasks by enhancing the capabilities of conversational models. Nonetheless, the application of ChatGPT in the recommendation domain has not been thoroughly investigated. In this paper, we employ ChatGPT as a general-purpose recommendation model to explore its potential for transferring extensive linguistic and world knowledge acquired from large-scale corpora to recommendation scenarios. Specifically, we design a set of prompts and evaluate ChatGPT's performance on five recommendation scenarios, including rating prediction, sequential recommendation, direct recommendation, explanation generation, and review summarization. Unlike traditional recommendation methods, we do not fine-tune ChatGPT during the entire evaluation process, relying only on the prompts themselves to convert recommendation tasks into natural language tasks. Further, we explore the use of few-shot prompting to inject interaction information that contains user potential interest to help ChatGPT better understand user needs and interests. Comprehensive experimental results on Amazon Beauty dataset show that ChatGPT has achieved promising results in certain tasks and is capable of reaching the baseline level in others. We conduct human evaluations on two explainability-oriented tasks to more accurately evaluate the quality of contents generated by different models. The human evaluations show ChatGPT can truly understand the provided information and generate clearer and more reasonable results. We hope that our study can inspire researchers to further explore the potential of language models like ChatGPT to improve recommendation performance and contribute to the advancement of the recommendation systems field. The prompts and codes are available in \url{https://github.com/williamliujl/LLMRec}.
\end{abstract}


\maketitle
\section{Introduction}
\label{intro}

As a crucial technique for addressing information overload and enhancing user experience, recommendation systems have witnessed significant advancements over the past decade and have been widely used in various web applications such as product recommendation \cite{sun2022revisiting, liu2022ecommerce,tsagkias2021challenges, xie2022decoupled}, video recommendation \cite{wei2019mmgcn,zhao2019recommending,papadamou2022just}, news recommendation \cite{wu2022feedrec,wu2020mind,wu2019npa}, music recommendation \cite{kowald2020unfairness,singh2022novel} and so on. In the meanwhile, with the development of deep learning, recommendation systems have gone through several stages. In early ages, collaborative filtering-based methods \cite{zhang2019privacy,bobadilla2020deep,bobadilla2020classification,rezaimehr2021survey} are primarily used to model the user's behavior patterns from the user-item interactions. Later on, with the introduction of user and item side information into recommendation systems, content-based recommendation \cite{musto2016learning,volkovs2017content,mittal2020smart,perez2021content,xie2023rethinking} and knowledge-based recommendation \cite{dong2020interactive,gazdar2020new,alamdari2020systematic,cena2021logical} have gained attention due to their ability to provide personalized recommendations. 
\par 
However, most traditional recommendation methods are task-specific. Therefore, specific data is required to train specific models for different tasks or application scenarios, which lack efficient generalization ability. To address this issue, researchers have shifted their focus towards implementing Pretrained Language Models (PLMs) in recommendation scenarios since PLMs have demonstrated impressive adaptability to improve the performance of downstream NLP tasks significantly~\cite{liu2023benchmarking}. To effectively convert user interaction data into text sequences, a variety of prompts \cite{zhang2021language} are designed to convert user interaction data into text sequences. Furthermore, P5 \cite{geng2022recommendation} and M6-Rec \cite{m6} focus on building a foundation model to support a wide range of recommendation tasks.
\par 
Recently, the emergence of ChatGPT has significantly advanced NLP tasks by enhancing the capabilities of conversational models, making it a valuable tool for businesses and organizations. Chataug et al. \cite{dai2023chataug} leverages ChatGPT to rephrase sentences for text data augmentation. Jiao et al. \cite{jiao2023chatgpt} finds the translation ability of ChatGPT performs competitively with commercial translation products on high-resource and low-resource languages. Bang et al. \cite{bang2023multitask} finds ChatGPT outperforms the previous state-of-the-art zero-shot model by a large margin in the sentiment analysis task. Nonetheless, the application of ChatGPT in the recommendation domain has not been thoroughly investigated, and whether ChatGPT can perform well on classical recommendation tasks remains an open question. Therefore, it is necessary to establish a benchmark to preliminarily evaluate and compare ChatGPT with traditional recommendation models, thereby providing valuable insights and facilitating further exploration of the potential of large-scale language models in recommendation systems.
\par
To bridge this research gap, in this paper, we directly employ ChatGPT as a general-purpose recommendation model that can handle various recommendation tasks, and attempt to explore whether the extensive linguistic and world knowledge acquired from large-scale corpora can be effectively transferred to recommendation scenarios. 
Our main contribution is the construction of a benchmark to track ChatGPT's performance in recommendation scenarios, and a comprehensive analysis and discussion of its strengths and limitations. Specifically, we design a set of prompts and evaluate ChatGPT's performance on five recommendation tasks, including rating prediction, sequential recommendation, direct recommendation, explanation generation, and review summarization. Unlike traditional recommendation methods, we do not fine-tune ChatGPT during the entire evaluation process, relying only on the prompts themselves to convert recommendation tasks into natural language tasks. Furthermore, we explore the use of few-shot prompting to inject interaction information that contains user potential interests to help ChatGPT better understand user needs and preferences.

Comprehensive experimental results on Amazon Beauty dataset reveal that, from the perspective of accuracy, ChatGPT performs well in rating prediction but poorly in sequential and direct recommendation tasks, achieving only similar performance levels to early baseline methods on certain metrics. 
On the other hand, while ChatGPT demonstrates poor performance in terms of objective evaluation metrics for explainable recommendation tasks such as explanation generation and review summarization, our additional human evaluations show that ChatGPT outperforms state-of-the-art methods. This highlights the limitations of using an objective evaluation approach to accurately reflect ChatGPT's true explainable recommendation capabilities.
Furthermore, despite ChatGPT's unsatisfactory performance in accuracy-based recommendation tasks, it is worth noting that ChatGPT has not been specifically trained on any recommendation data. Thus, there is still significant potential for improvement in future research by incorporating more relevant training data and techniques.
We believe that our benchmark not only sheds light on ChatGPT's recommendation capabilities but also provides a valuable starting point for researchers to better understand the advantages and shortcomings of ChatGPT in recommendation tasks. Moreover, we hope that our study can inspire researchers to design new methods that leverage the strengths of language models like ChatGPT to improve recommendation performance, and contribute to the advancement of the recommendation systems field.

\par

\section{Related Work}
\label{intro}
\par  

\subsection{Large Language Models and ChatGPT}
Language Models (LMs) are a fundamental component of natural language processing (NLP) and have been the focus of research for several decades. Recently, the emergence of large-scale LMs has led to significant progress in NLP tasks such as machine translation\cite{chen2018best,aharoni2019massively,zeng2022greenplm}, summarization\cite{2017Get,2019Fine}, and dialogue generation\cite{2016A,2016Towards}. 

Large Language Models (LLMs) are a subclass of LMs that leverage massive amounts of data and computational resources to achieve state-of-the-art performance on a wide range of NLP tasks. The history of LLMs can be traced back to the early work on neural networks and language modeling. \cite{2003Journal} introduced neural language models that learned to predict the next word in a sentence given the previous words. Later, the development of recurrent neural networks (RNNs) and long short-term memory (LSTM) networks further improved the ability of models to capture long-term dependencies in language\cite{1997Long}. However, traditional neural language models still struggled with capturing the rich semantic and contextual relationships present in natural language. The introduction of the Transformer architecture by \cite{2017Attention} was a major breakthrough in this area. The Transformer model utilizes self-attention mechanisms to capture the relationships between all elements in a sequence simultaneously, allowing for more comprehensive contextual understanding. This architecture has been used as the backbone of many successful LLMs, including BERT\cite{bert}, GPT-2\cite{radford2019language}, and XLNet\cite{2019XLNet}.

ChatGPT\cite{gpt4} is a state-of-the-art dialogue system developed by OpenAI in 2022. It is a state-of-the-art natural language processing (NLP) model that has been widely used in various vertical domains, such as text generation and dialogue systems. In text generation, ChatGPT has shown impressive results in generating coherent and diverse text, surpassing the performance of previous models \cite{2020Language}. In dialogue systems, it has been used for task-oriented and open-domain conversations, achieving state-of-the-art performance in both settings \cite{2019DialoGPT}. 
Although the value of ChatGPT has been validated in various fields, whether it can still be effective in the recommendation domain remains an under-explored topic, which motivates us to construct such a benchmark to gain insights into the potential of large language models for recommendation systems.

\label{intro}
\par  
\begin{figure}[t]
\centering
\includegraphics[width=8.5cm]{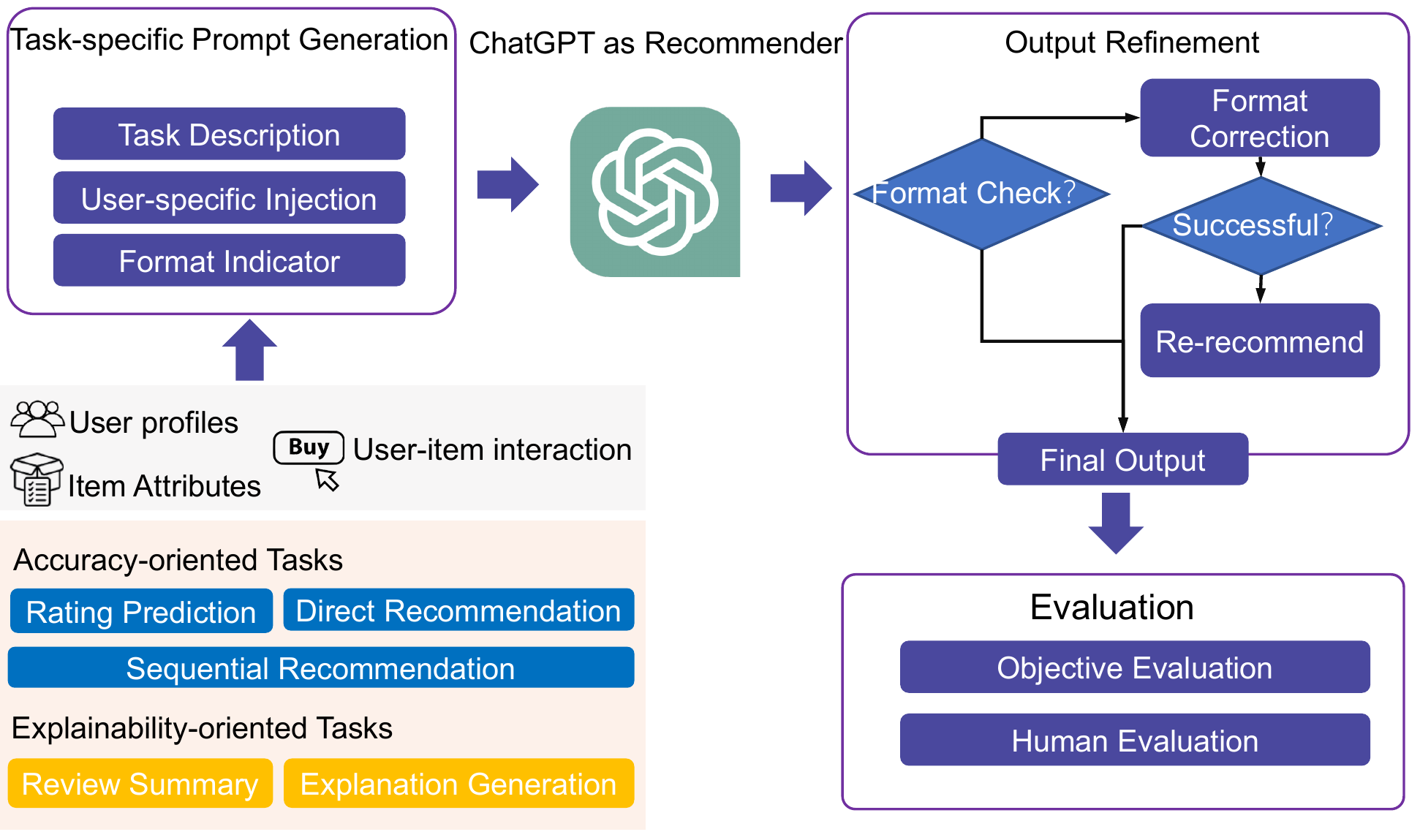}
\caption{Workflow of utilizing ChatGPT to perform five recommendation tasks and evaluate its recommendation performance.}
\label{fig:workflow}
\end{figure}

\subsection{Language Model for Recommendation}
Language Models (LMs), such as BERT~\cite{bert} and GPT~\cite{gpt4}, have demonstrated impressive adaptability to improve the performance of downstream NLP tasks significantly, thanks to extensive linguistic and world knowledge learned from large-scale corpora. 
Inspired by these achievements, an increasing amount of attention is being paid to the application of LMs in recommender scenarios, yielding several recent breakthroughs in this field.
For instance, LMRecSys~\cite{zhang2021language} utilizes prompts to reconstitute some recommendation tasks as multi-token cloze tasks, aiming to address zero-shot and data efficiency issues. P5~\cite{geng2022recommendation} is the first attempt to integrate different recommendation tasks within a shared conditional language generation framework (i.e., T5~\cite{t5}). To effectively convert user interaction data into text sequences, a variety of prompts are designed to accommodate the specific characteristics of each recommendation task. Similarly, M6-Rec~\cite{m6} focuses on building a foundation model to support a wide range of recommendation tasks, including retrieval, ranking, and explanation generation, etc. Notably, the authors also provide practical solutions for model deployment in real-world settings. Chat-REC~\cite{chat-rec}, a concurrent work closely related to our study, leverages ChatGPT as an interface for conversational recommendations, thereby augmenting the performance of existing recommender models and rendering the recommendation process more interactive and explainable.

Different from Chat-REC, our work is inspired by P5 and treats ChatGPT as a self-contained recommendation system that does not rely on any external systems.  Based on this, we conduct a thorough evaluation and comparison of its performance on classic recommendation tasks including sequential recommendation, rating prediction, etc. By doing so, we hope our analysis can offer valuable insights for researchers to delve deeper into the potential of large-scale language models in the domain of recommendation.
\section{Recommendation with ChatGPT}

\begin{figure*}[htb]
\centering
\includegraphics[width=17cm]{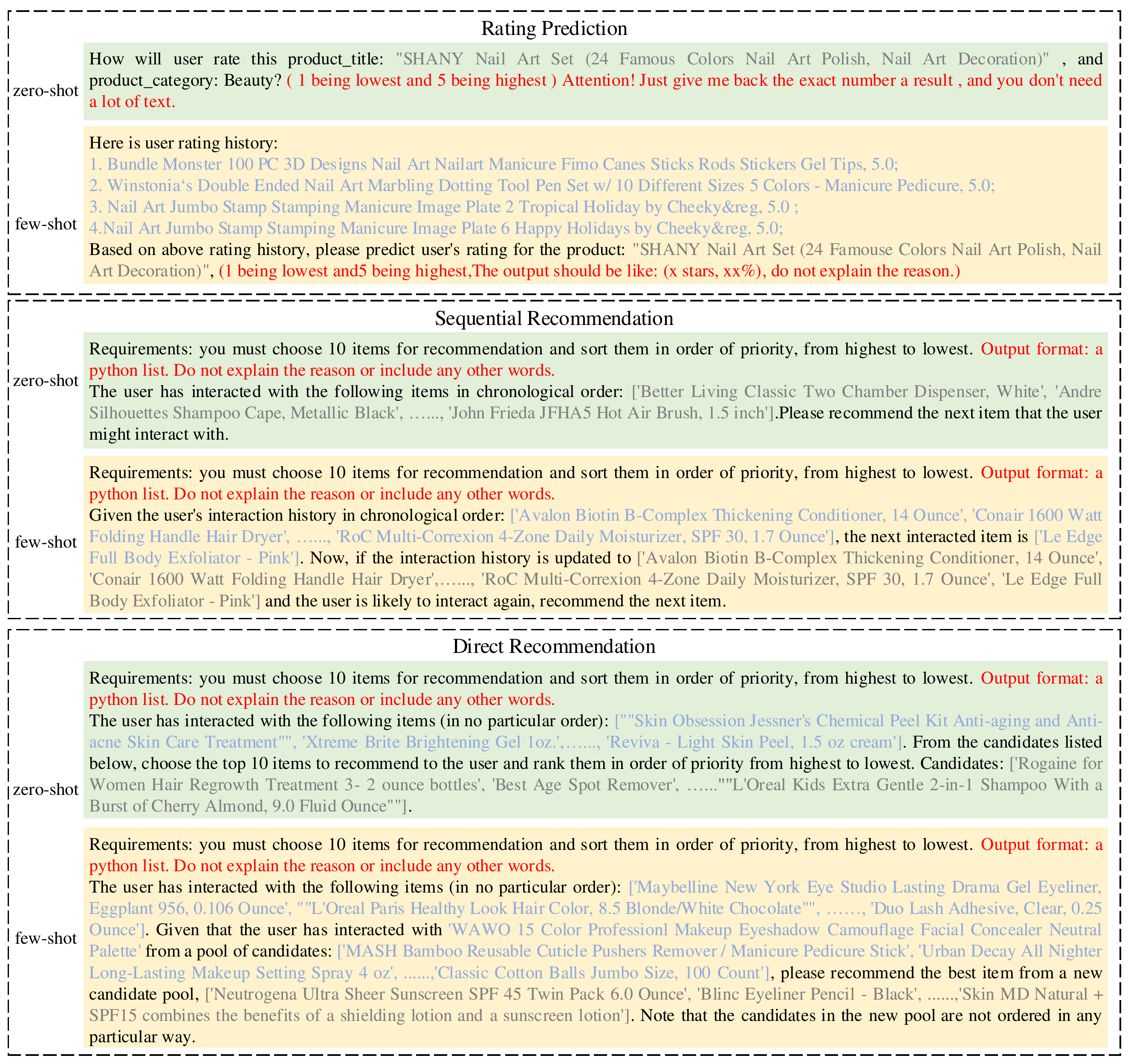}
\caption{Example prompts of accuracy-based tasks on \textit{Beauty} dataset. The black texts represent the description of the task, the \textcolor[RGB]{255,0,0}{red} texts indicate the format requirements, the \textcolor[RGB]{144,170,219}{blue} texts represent user historical information or few-shot information, and the \textcolor[RGB]{127,127,127}{gray} texts indicate the current input.}
\label{fig:accuracy-based}
\end{figure*}

The workflow of using ChatGPT to complete recommendation tasks is illustrated in Fig.\ref{fig:workflow}, which consists of three steps. First, different prompts are constructed based on the specific characteristics of the recommendation tasks (Section 3.1). Second, these prompts are used as inputs for ChatGPT, which generates the recommendation results according to the requirements specified in the prompts. Finally, the output from ChatGPT is checked and refined by the refinement module, and the refined results are returned to the user as the final recommendation results (Section 3.2).
\subsection{Task-specific Prompt Construction}
In this section, we investigate the recommendation capability of ChatGPT by designing prompts tailored to different tasks. Each prompt comprises three parts: task description, behavior injection, and format indicator. The task description is utilized to adapt recommendation tasks to natural language processing tasks. The behavior injection is designed to assess the impact of few-shot prompting, which incorporates user-item interaction to aid ChatGPT in capturing user preferences and needs more effectively. The format indicator serves to constrain the output format, making the recommendation results more comprehensible and assessable.
\subsubsection{Rating Prediction}
Rating prediction is a crucial task in recommendation systems that aims to predict the ratings that a user would give to a particular item. This task is essential in personalizing recommendations for users and improving the overall user experience. Some recent advancements in this field include the use of deep learning models\cite{2017Neural}, and the use of matrix factorization techniques\cite{2009Matrix}, which are effective in dealing with the sparsity problem in recommendation systems. In line with the innovative recommendation paradigm of the LLM, we conducted experiments on a rating task that involved formulating two unique prompt types to elicit the results. We provide some sample prompts in Fig.\ref{fig:accuracy-based}.

\begin{figure*}[htb]
\centering
\includegraphics[width=17cm]{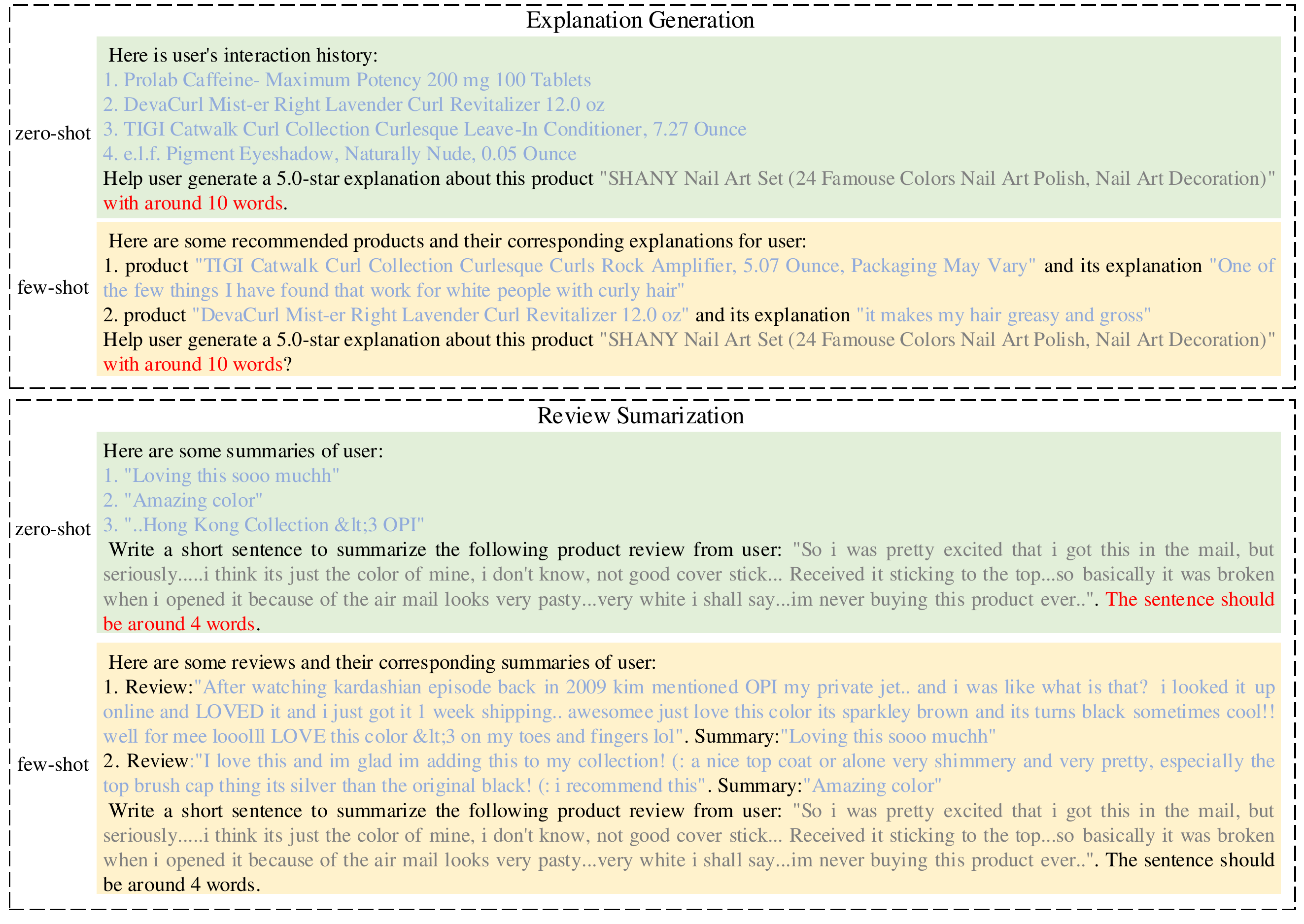}
\caption{Example prompts of explainability-oriented tasks on \textit{Beauty} dataset.  The black texts represent the description of the task, the \textcolor[RGB]{255,0,0}{red} texts indicate the format requirements, the \textcolor[RGB]{144,170,219}{blue} texts represent user historical information or few-shot information, and the \textcolor[RGB]{127,127,127}{gray} texts indicate the current input.}
\label{fig:explainability-oriented}
\end{figure*}

\subsubsection{Sequential Recommendation}
Sequential recommendation is a subfield of recommender systems that aims to predict a user's next item or action based on their past sequential behavior. It has received increasing attention in recent years due to its potential applications in various domains, such as e-commerce, online advertising, and music recommendation. In sequential recommendation, researchers have proposed various methods, including recurrent neural networks\cite{2015A}, contrastive learning\cite{zhou2023equivariant}, and attention-based models\cite{2017Attention,zhou2023attention}, for capturing the temporal dependencies and patterns in user-item interactions. We have devised three distinct prompt formats for the sequential recommendation task family. These include: 1) direct prediction of the user's next item based on their interaction history, 2) selection of a possible next item from a list of candidates, where only one item is positive and based on the user's interaction history, and 3) prediction of whether a specific item will be the next one interacted with by the user, using their previous interaction history as a basis. These prompt formats have been designed to enhance the accuracy and effectiveness of sequential recommendations, and are grounded in rigorous academic principles. Examples of these prompts can be seen in Fig.\ref{fig:accuracy-based}.
\subsubsection{Direct Recommendation}
Direct Recommendation, also known as explicit feedback recommendation or rating-based recommendation, is a type of recommendation system that relies on explicit feedback from users in the form of ratings or reviews. Unlike other recommendation systems that rely on implicit feedback, such as user behavior or purchase history, direct recommendation systems are able to provide more personalized and accurate recommendations by taking into account the explicit preferences of users. For this task, we develop the item selection prompt that selects the most appropriate item from a list of potential candidates. These prompt formats are based on rigorous academic principles and aim to optimize the accuracy and relevance of recommendations. Examples of these prompts can be seen in Fig.\ref{fig:accuracy-based}.

\subsubsection{Explanation Generation}
Explanation generation refers to providing users or system designers with explanations to clarify why such items are recommended. In this way, it enhances the transparency, persuasiveness, effectiveness, trustworthiness, and user satisfaction of recommendation systems. Furthermore, it facilitates system designers in diagnosing, debugging, and refining the recommendation algorithm.
Large language models such as ChatGPT can use the vast amount of knowledge they contain to learn the user's interests through their historical interaction records and provide reasonable explanations for their behavior. 
Specifically, We ask ChatGPT model to generate a textual explanation to justify a user's preference towards a selected item as shown in Fig.\ref{fig:explainability-oriented}. For each category, additional auxiliary information such as the hint word and the star rating could be included.

\subsubsection{Review Summarization}
Automatic generation of summaries is becoming increasingly important in Natural Language Processing, as the demand for concise and easily comprehensible content continues to grow. Similar to the explanation generation task, we create two types of prompts: zero/few-shot prompts, and provide some example prompts in Fig.\ref{fig:explainability-oriented}.




\subsection{Output Refinement}
To ensure the diversity of generated results, ChatGPT incorporates a degree of randomness into its response generation process, which may result in different responses for the same input. However, when using ChatGPT for recommendation, this randomness can sometimes cause difficulties in evaluating the recommended items. While the format indicator in the prompt construction can partially alleviate this issue, in practical usage, it still cannot guarantee the anticipated output format. Therefore, we devise output refinement module to check the format of ChatGPT's output. If the output passes the format check, it is directly used as the final output. If not, it is modified based on pre-defined rules. If the format correction is successful, the corrected result is used as the final output. If not, the corresponding prompt is fed into ChatGPT for a re-recommendation until the format requirements are met. It is worth noting that different tasks have different output format requirements when evaluating ChatGPT. For example, for rating prediction, only a specific score is needed, whereas for sequential or direct recommendation, a list of recommended items is required. Particularly for sequence recommendation, it is challenging to feed all the items in the dataset to ChatGPT at once. As a result, ChatGPT's output may not correctly match the item set in the dataset. To address this issue, we introduce a text matching method based on similarity in the correction process to map ChatGPT's predictions back to the original dataset. Although this method may not perfectly reflect ChatGPT's ability, it can still indirectly demonstrate its potential in sequential recommendation.

\section{Evaluation}
To evaluate ChatGPT, we conduct extensive experiments on the real-world Amazon dataset. Through the performance comparison with various representative methods and ablation studies on different tasks, we aim to answer the following research questions:
\begin{itemize}
    \item \textbf{RQ1}: How does ChatGPT perform as compared with the state-of-the-art baseline models?
    \item \textbf{RQ2}: What is the impact of few-shot prompting on performance?
    \item \textbf{RQ3}: How do we design the human evaluation to assess explanation generation and summarization tasks?
\end{itemize}

\subsection{Experimental Setup}
\subsubsection{Datasets}
We conduct numerical and human evaluations on the real-world Amazon recommendation dataset. The Amazon dataset contains the customer review text with accompanying metadata on 29 categories of products. This paper focuses on evaluating the \textit{Beauty} category.

\subsubsection{Metrics}
In numerical evaluations, we employ Root Mean Square Error (RMSE) and Mean Absolute Error (MAE) for rating prediction. And we adopt top-\textit{k} Hit Ratio (HR@\textit{k}), top-\textit{k} Normalized Discounted Cumulative Gain (NDCG@\textit{k}) for sequential recommendation and direct recommendation which are widely used in related works \cite{geng2022recommendation,zhou2020s3}. Specifically, we report results on HR@\{1,5,10\}, NCGG@\{5,10\} for evaluation. Besides, \textit{n}-gram Bilingual Evaluation Understudy (BLEU-\textit{n}) and 
\textit{n}-gram Recall-Roiented Understudy for Gising Evaluation (ROUGE-\textit{n}) are used to evaluate the explanation generation and review summarization tasks. In human evaluations, we have designed and deployed a crowdsourcing task to assess the qualities of the generated explanations and review summaries. Through this task, we aim to accurately evaluate the effectiveness of the content by gathering feedback from a diverse range of human evaluators.

\subsubsection{Implementation Details}
In order to verify that we can directly apply the knowledge learned by ChatGPT to recommendation scenarios without the need for a large amount of task-specific data for training, we apply \textit{gpt-3.5-turbo} to conduct few-shot and zero-shot experiments for the five tasks mentioned above. We collect \textit{n} items that users have interacted with and \textit{k} shots of historical records to enable ChatGPT to learn users' interests implicitly. In this experiment, we use the titles of the items as meta information, and set $n=10$ and $k=3$ due to the limitation of a maximum context length of 4096 tokens in ChatGPT. 
We randomly sample 100 records from the test set proposed by P5 \cite{geng2022recommendation} for evaluation. For direct recommendation, we set the number of negative samples to 99, thus forming a candidate list of length 100 with one positive item. Also, due to the addition of the candidate pool in the request, we set the number of shots to 1. 
For sequential recommendation, we input the user's historical interacted items in order and let ChatGPT predict the title of the next item that the user might interact with, and use BERT\cite{bert} to calculate the vector of the predicted title and compute the similarity between the predicted title vector and the title vectors of all items, and select the item with the highest similarity as the predicted item. 
For human evaluation on explanation generation and review summarization, we sample some results of different methods for each task, and each result will be scored and ranked by three human evaluators. After obtaining the manually annotated results, we will calculate the average top1 ratio and average ranking position of different methods to measure their generation performance.

\subsection{Baselines for multiple tasks}
Following P5 \cite{geng2022recommendation}, we gather a range of approaches that are representative of various tasks. For rating prediction, we employ MF \cite{koren2009matrix} and MLP \cite{cheng2016wide} as our baselines, both evaluated using mean square root loss. For direct recommendation, we use BPR-MF \cite{rendle2012bpr}, BPR-MLP \cite{cheng2016wide} and SimpleX \cite{mao2021simplex} as baselines. For sequential recommendation, we adopt Caser \cite{tang2018personalized}, HGN \cite{ma2019hierarchical}, GRU4Rec \cite{hidasi2015session}, BERT4Rec \cite{sun2019bert4rec}, FDSA \cite{zhang2019feature}, SASRec \cite{kang2018self} and ${\rm S}^3$-Rec \cite{zhou2020s3} as baselines for comparison. For explanation generation, we utilize Attn2Seq \cite{dong2017learning}, NRT \cite{li2017neural} and PETER \cite{li2021personalized} as baselines. For review summarization, we adopt pretrained T0 \cite{sanh2021multitask} and GPT-2 \cite{radford2019language} as baselines. For more details, you can refer to P5 \cite{geng2022recommendation} or relevant articles.

\begin{table}[t!]
\centering
\footnotesize
\caption{Performance comparison on rating prediction.}
\begin{adjustbox}{width=0.65\linewidth}
\begin{tabular}{ccc}
\toprule
\multirow{2.5}{*}{Methods} & \multicolumn{2}{c}{\textbf{Beauty}} \\
\cmidrule(lr){2-3}
 & RMSE  & MAE  \\
\cmidrule{1-3}
MF    &  1.1973 & 0.9461   \\
MLP    & 1.3078  & 0.9597  \\
ChatGPT(zero-shot)    &  1.4059   & 1.1861   \\
ChatGPT(few-shot)   & \bf 1.0751  & \bf 0.6977    \\
\bottomrule
\end{tabular}
\end{adjustbox}
\label{tab:rating}
\vspace{-10pt}
\end{table}

\subsection{Performance Comparison on 5 Tasks (RQ1\&2)}

\subsubsection{Rating prediction}
To evaluate the rating prediction performance of ChatGPT, zero-shot and few-shot prompts were employed, and the results obtained from the Beauty dataset were summarized in Tab.\ref{tab:rating}. The results indicate that, for the seen category on the Beauty dataset, few-shot prompts outperform MF and MLP in terms of both MAE and RMSE. These results provide evidence supporting the feasibility of utilizing a conditional text generation framework for rating prediction.

\subsubsection{Sequential recommendation}
To assess the sequential recommendation capability of ChatGPT, we conducted both zero-shot and few-shot experiments, the results of which are shown in Tab.\ref{tab:sequential}. We found that, compared to the baselines, ChatGPT's performance in the zero-shot prompting setup is considerably inferior, with all metrics being significantly lower than the baselines. However, under the few-shot prompting setup, while there is a relative improvement in performance, such as NDCG@5 surpassing GRU4Rec, ChatGPT is still generally outperformed by classical sequential recommendation methods in most cases. There are possibly two main reasons contributing to this outcome: First, during the prompting design process, all items are represented by their titles. Although this approach can alleviate the cold-start problem to some extent, it may cause ChatGPT to focus more on semantic similarity rather than the transition relationships between items, which are crucial for effective recommendations. Second, due to the length constraint of the prompts, it is not possible to input all items from the item set into ChatGPT. This leads to ChatGPT lacking constraints in predicting the title of the next item, resulting in generating item titles that do not exist in the dataset. Although it is possible to map these predicted titles to existing titles in the dataset through semantic similarity matching, our experiments show that this mapping does not result in significant gains. Therefore, for sequential recommendation tasks, merely employing ChatGPT is not a suitable choice. Further exploration is needed to introduce more guidance and constraints to help ChatGPT accurately capture historical interests and make reasonable recommendations within a limited scope.
\begin{table}
\centering
\caption{Performance comparison on sequential recommendation.}
\label{tab:sequential}
\begin{adjustbox}{width=0.97\linewidth}
\begin{tabular}{ccccc} 
\toprule
\multirow{2.5}{*}{Methods}  & \multicolumn{4}{c}{\textbf{Beauty}}  \\ 
\cmidrule(l){2-5}
          & HR@5   & NDCG@5 & HR@10  & NDCG@10   \\ 
\midrule
Caser     & 0.0205 & 0.0131 & 0.0347 & 0.0176    \\
HGN       & 0.0325 & 0.0206 & 0.0512 & 0.0266    \\
GRU4Rec   & 0.0164 & 0.0099 & 0.0283 & 0.0137    \\
BERT4Rec  & 0.0203 & 0.0124 & 0.0347 & 0.0170    \\
FDSA      & 0.0267 & 0.0163 & 0.0407 & 0.0208    \\
SASRec    & 0.0387 & 0.0249 & 0.0605 & 0.0318    \\
S$^3$-Rec & 0.0387 & 0.0244 & 0.0647 & 0.0327    \\
P5-B      & \textbf{0.0493} & \textbf{0.0367} & \textbf{0.0645} & \textbf{0.0416}    \\
ChatGPT(zero-shot)   & 0.0000  & 0.0000  &  0.0000 &  0.0000   \\
ChatGPT(few-shot)   & 0.0135 & 0.0135 & 0.0135 & 0.0135\\
\bottomrule
\end{tabular}
\end{adjustbox}
\end{table}
\subsubsection{Direct recommendation}
Tab.\ref{tab:direct} illustrates the performance of ChatGPT on the direct recommendation task. Unlike the sequential recommendation setup, direct recommendation requires the recommendation model to select the most relevant item for the user from a limited-sized item pool. We observed that, when using zero-shot prompting, the recommendation performance is significantly inferior to supervised recommendation models. This can be attributed to the insufficient information provided to ChatGPT, resulting in an inability to capture user interests and generate more random recommendations. While few-shot prompting can improve ChatGPT's recommendation performance by providing some of the user's historical preferences, it still fails to surpass the baseline performance.

It is worth noting that we discovered during the experiments that the construction of the item pool, specifically whether the item pool is shuffled or not, has a considerable impact on the direct recommendation performance. In an extreme scenario where the ground truth item is placed at the first position in the item pool, we found that the evaluation metrics were approximately ten times higher than when the item pool was shuffled. This finding suggests that ChatGPT exhibits a positional bias for the input item pool within the prompt, tending to consider items towards the beginning of the pool as more important, and thus more likely to be recommended. This additional bias introduced by the language model renders using ChatGPT for direct recommendation a challenging endeavor.
\begin{table}
\centering
\caption{Performance comparison on direct recommendation.}
\label{tab:direct}
\begin{adjustbox}{width=0.97\linewidth}
\begin{tabular}{ccccc} 
\toprule
\multirow{2.5}{*}{Methods}             & \multicolumn{4}{c}{\textbf{Beauty}}                        \\ 
\cmidrule(l){2-5}
\multicolumn{1}{l}{} & HR@5           & NDCG@5         & HR@10  & NDCG@10         \\ 
\cmidrule{1-5}
BPR-MF               & 0.1426         & 0.0857         & 0.2573 & 0.1224          \\
BPR-MLP              & 0.1392         & 0.0848         & 0.2542 & 0.1215          \\
SimpleX              & \textbf{0.2247} & \textbf{0.1441} & \textbf{0.3090} & \textbf{0.1711}  \\ 
P5-B                 & 0.1564         & 0.1096         & 0.2300 & 0.1332          \\
ChatGPT(zero-shot) &    0.0217 & 0.0111 & 0.0652 & 0.0252    \\
ChatGPT(few-shot)     & 0.0349 &     0.0216 &      0.0930 &  0.0398                \\
\bottomrule
\end{tabular}
\end{adjustbox}
\end{table}

\subsubsection{Explanation Generation}
\begin{figure*}[htb]
\centering
\includegraphics[width=16cm]{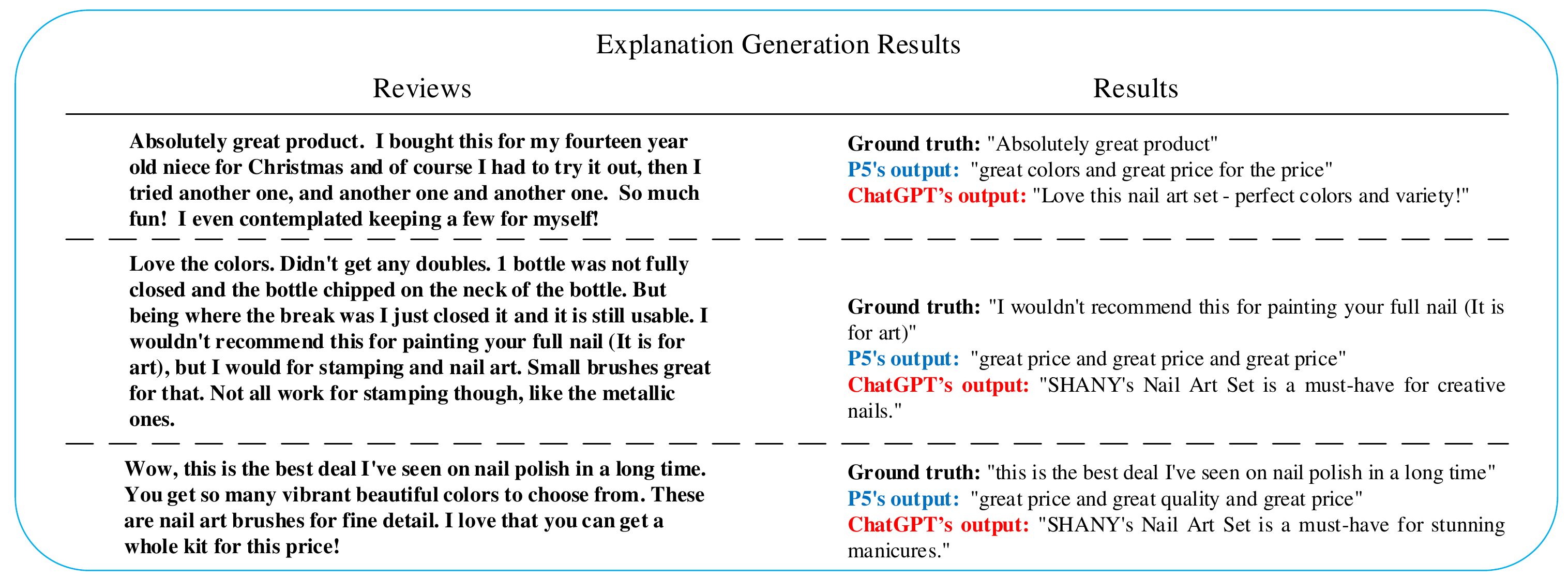}
\caption{Example explanation results of different models on \textit{Beauty} dataset.}
\label{fig:explanation_results}
\end{figure*}

In Tab.\ref{tab:explanation}, both zero-shot and few-shot prompts are used to evaluate ChatGPT's performance on explanation generation. From the metrics perspective, the P5 model has a better performance. As language models, P5 and ChatGPT have different design goals and application scenarios. P5 aims to generate explanatory language similar to known texts. Therefore, P5 focuses on learning text structure and grammar rules during training, making the generated results more standardized, as shown in Fig.\ref{fig:explanation_results}. In contrast, ChatGPT focuses more on language interaction and diversity. Its application scenario is usually to simulate human conversation, so it needs to consider multiple factors such as context, emotion, and logic when generating text to better express human thinking and language habits. This design is bound to make the text generated by ChatGPT more diverse and creative. Besides, P5 is fine-tuned on \textit{Beauty} dataset while ChatGPT is utilized in a zero-shot or few-shot experimental setting. Therefore, it is understandable that ChatGPT may not perform as well as P5 in metrics. Hence, we introduce human evaluation to better measure the performance of different models in generating content.

\begin{table}[t]
\centering
\caption{Performance comparison on explanation generation (\%).}
\vspace{-10pt}
\begin{adjustbox}{width=0.97\linewidth}
\begin{tabular}{ccccc}
\toprule
\multirow{2.5}{*}{Methods}  & \multicolumn{4}{c}{\textbf{Beauty}}  \\
\cmidrule(lr){2-5}
 & BLUE4  & ROUGE1 & ROUGE2  & ROUGEL   \\
\cmidrule{1-5}
Attn2Seq    & 0.7889 & 12.6590 & 1.6820 & 9.7481   \\
NRT    & 0.8295  & 12.7815  & 1.8543  & 9.9477   \\
PETER   & 1.1541  & 14.8497  & 2.1413  & 11.4143  \\
P5-B   & 0.9742  & 16.4530  & 1.8858  & 11.8765  \\
PETER+   & \bf \textbf{3.2606}  & \textbf{25.5541}  &  \textbf{5.9668}  &  \textbf{19.7168}  \\
ChatGPT(zero-shot)  &0.0000 & 8.5992 & 0.6995 & 4.7564\\
ChatGPT(few-shot)  &1.1967 & 11.4103 & 2.5675 & 5.9119\\
\bottomrule
\end{tabular}
\end{adjustbox}
\vspace{-5pt}
\label{tab:explanation}
\end{table}

\subsubsection{Review summarization}
\begin{figure*}[htb]
\centering
\includegraphics[width=16cm]{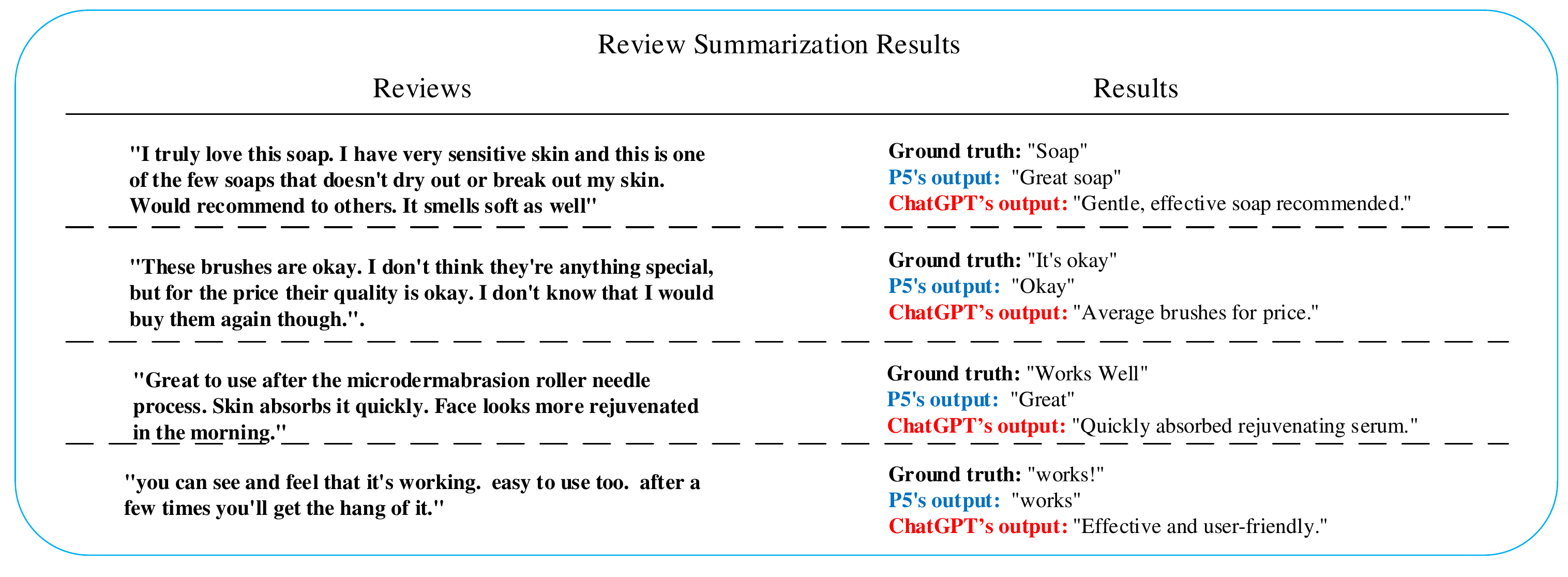}
\caption{Example summarization results of different models on \textit{Beauty} dataset.}
\label{fig:summarization_results}
\end{figure*}

We conduct zero-shot and few-shot experiments to evaluate ChatGPT's ability on review summarization, as shown in Tab.\ref{tab:summarization}. Similar to the explanation generation task, ChatGPT does not have an advantage in metrics evaluation. However, although the summary result of P5 has extracted some keywords, it has ignored relevant information from the entire review. In contrast, ChatGPT can generate more effective and meaningful summaries by deeply understanding and summarizing the reviews. As shown in Fig.\ref{fig:summarization_results}. Hence, we also conduct human evaluation in this task.

\begin{table}[t]
\centering
\caption{Performance comparison on review summarization (\%).}
\vspace{-10pt}
\begin{adjustbox}{width=0.97\linewidth}
\begin{tabular}{ccccc}
\toprule
\multirow{2.5}{*}{Methods}  & \multicolumn{4}{c}{\textbf{Beauty}}  \\
\cmidrule(lr){2-5}
 & BLUE4  & ROUGE1 & ROUGE2  & ROUGEL   \\
\cmidrule{1-5}
T0   &  1.2871  & 1.2750  &  0.3904  &  0.9592  \\
GPT-2    &  0.5879 &  3.3844 & 0.6756  & 1.3956 \\
P5-B   &  \textbf{2.1225} &  \textbf{8.4205} & \textbf{1.6676}  & \textbf{7.5476} \\
ChatGPT(zero-shot)  &0.0000 & 3.8246 & 0.2857 & 3.1344\\
ChatGPT(few-shot)  &0.0000 & 2.7822 & 0.0000 & 2.4328\\
\bottomrule
\end{tabular}
\end{adjustbox}
\vspace{-5pt}
\label{tab:summarization}
\end{table}




\subsection{Human Evaluation (RQ3)}

As shown in the experiments above, we conducted numerical evaluations on the explanation generation and review summarization tasks using the test set constructed by P5. However, the ground-truth explanations generated by P5 are not truly accurate because P5 extracts sentences from views commenting on one or more item feature words as users' explanations about their preferences. In that case,  we designed human evaluations to better assess the performance of ChatGPT.
Specifically, we randomly sample 20 prompts for explanation generation and 97 prompts for review summarization from the \textit{Beauty} dataset and display every generated explanation or summary to several evaluators. The evaluators rank the results generated by ChatGPT, baseline, and ground truth for assessment. avg\_top1\_ration represents the proportion in which the prompt ranked first among the prompts. avg\_position denotes the average position of sorting for each prompt. 

\begin{table}
\centering
\caption{Human evaluation for explanation generation on \textit{Beauty} dataset.}
\label{tab:human_evaluation_for_explanation}
\begin{adjustbox}{width=0.97\linewidth}
\begin{tabular}{ccccccc} 
\toprule
\multirow{2.5}{*}{Methods} & \multicolumn{4}{c}{\textbf{Evaluators}}                        & \multicolumn{1}{c}{\multirow{2.5}{*}{avg\_top1\_ration}} & \multicolumn{1}{c}{\multirow{2.5}{*}{avg\_position}}  \\ 
\cmidrule(r){2-5}
                         & Eva\_1 & Eva\_2 & Eva\_3 & Eva\_4 & \multicolumn{1}{c}{}   & \multicolumn{1}{c}{}       \\ 
\midrule
Ground truth                    & 25.0\% & 45.0\% &45.0\% & 50.0\% & 38.0\% &  1.83    \\
P5                      & 0.0\% & 0.0\% &0.0\%&0.0\%  & 0.0\%  & 2.71    \\
ChatGPT(zero-shot)                  & 75.0\% & 55.0\% &55.0\% & 50.0\%  & \textbf{62.0\%}  &  \textbf{1.46}    \\
\bottomrule
\end{tabular}
\end{adjustbox}
\end{table}

\begin{table}
\centering
\caption{Human evaluation for review summarization on \textit{Beauty} dataset.}
\label{tab:human_evaluation_for_summarization}
\begin{adjustbox}{width=0.97\linewidth}
\begin{tabular}{cccccccc} 
\toprule
\multirow{2.5}{*}{Methods} & \multicolumn{5}{c}{\textbf{Evaluators}}                        & \multicolumn{1}{c}{\multirow{2.5}{*}{avg\_top1\_ration}} & \multicolumn{1}{c}{\multirow{2.5}{*}{avg\_position}}  \\ 
\cmidrule(r){2-6}
                         & Eva\_1 & Eva\_2 & Eva\_3 & Eva\_4 & \multicolumn{1}{c}{Eva\_5} & \multicolumn{1}{c}{}                     & \multicolumn{1}{c}{}                                \\ 
\midrule
Ground truth                    & 12.5\% & 10.6\% &8.7\% & 17.3\% & 22.1\% & 14.2\% &  2.91    \\
P5                      & 5.8\% & 0.0\% &5.7\%&11.5\% & 19.2\% & 8.5\% & 3.16    \\
ChatGPT(zero-shot)                  & 46.2\% & 37.5\% &36.5\% & 45.2\% & 23.1\% & 37.7\% &  \textbf{1.90}    \\
ChatGPT(few-shot)                  & 35.6\% & 51.9\% &49.0\% & 26.0\% & 35.6\% & \textbf{39.6\%} &  2.01    \\
\bottomrule
\end{tabular}
\end{adjustbox}
\end{table}

\par
For explanation generation task, as shown in Tab.\ref{tab:human_evaluation_for_explanation}, the results of the four manual annotators have a certain degree of subjectivity, but the score distribution is relatively consistent, with a general consensus that the explanations generated by ChatGPT are clearer and more reasonable, even better than the ground truth. Meanwhile, P5's performance is the worst, with explanations tending towards a generic style and sentences that are not fluent. We can also draw the same conclusion from the examples in Tab.\ref{fig:explanation_results}. For review summarization task, we can find in Fig.\ref{fig:summarization_results} that the contents summarized in P5 are too general and do not extract useful information. However, ChatGPT can truly understand the reviews and provide accurate summaries, rather than simply extracting a few keywords from the reviews. As shown in Tab.\ref{tab:human_evaluation_for_summarization}, all annotators unanimously agree that ChatGPT has the best performance, surpassing ground truth and P5 by a large margin.

\section{Conclusion and future work}

 In this paper, we construct a benchmark to evaluate ChatGPT's performance in recommendation tasks and compare it with traditional recommendation models. The experimental results show that ChatGPT performs well in rating prediction but poorly in sequential and direct recommendation tasks, indicating the need for further exploration and improvement. Despite its limitations, ChatGPT outperforms state-of-the-art methods in terms of human evaluation for explainable recommendation tasks, highlighting its potential to generate explanations and summaries. We believe that our study provides valuable insights into the strengths and limitations of ChatGPT in recommendation systems, and we hope that it can inspire future research to explore the use of large language models to enhance recommendation performance. Moving forward, we plan to investigate better ways to incorporate user interaction data into large language models and bridge the semantic gap between language and user interests.
\appendix

\newpage

\balance
\bibliographystyle{ACM-Reference-Format}
\bibliography{references}

\end{document}